\documentclass[12pt]{article}
\usepackage{graphics,cite,amssymb,epsfig,float}
\usepackage[usenames,dvips]{color}
\usepackage{amsmath}

\def\lsim{\;\raise0.3ex\hbox{$<$\kern-0.75em\raise-1.1ex\hbox{$\sim$}}\;}
\def\gsim{\;\raise0.3ex\hbox{$>$\kern-0.75em\raise-1.1ex\hbox{$\sim$}}\;}
\def\beq{\begin{equation}}   \def\eeq{\end{equation}}
\def\ba{\begin{array}}       \def\ea{\end{array}}
\def\bea{\begin{eqnarray}}   \def\eea{\end{eqnarray}}
\def\nn{\nonumber}
\def\k{\kappa}
\def\l{\lambda}
\def\b{\beta}
\def\t{\theta}

\begin{document}

\begin{titlepage}
\begin{flushright}
LPT Orsay 10-53 \\ 
\end{flushright}

\begin{center}
\vspace{1cm}
{\Large\bf Light dark matter in the NMSSM: upper bounds on direct
detection cross sections} \\
\vspace{2cm}

{\bf{Debottam Das and Ulrich Ellwanger}}
\vspace{1cm}\\
\it  Laboratoire de Physique Th\'eorique, UMR 8627, CNRS and
Universit\'e de Paris--Sud,\\
\it B\^at. 210, 91405 Orsay, France \\

\end{center}

\vspace{1cm}

\begin{abstract}
In the Next-to-Minimal Supersymmetric Standard Model, a bino-like LSP
can be as light as a few GeV and satisfy WMAP constraints on the dark
matter relic density in the presence of a light CP-odd Higgs scalar. We
study upper bounds on the direct detection cross sections for such a
light LSP in the mass range $2-20$~GeV in the NMSSM, respecting all
constraints from B-physics and LEP. The OPAL constraints on $e^+ e^- \to
\chi^0_1 \chi^0_i$ ($i > 1$) play an important r\^ole and are discussed
in some detail. The resulting upper bounds on the spin-independent and
spin-dependent nucleon cross sections are $\sim 10^{-42}$~cm$^{-2}$ and
$\sim 4\times 10^{-40}$~cm$^{-2}$, respectively. Hence the upper bound
on the spin-independent cross section is below the DAMA and CoGeNT
regions, but could be compatible with the two events observed by
CDMS-II.
\end{abstract}

\end{titlepage}

\section{Introduction}

The DAMA \cite{Bernabei:2008yi} and CoGeNT \cite{Aalseth:2010vx} dark
matter detection experiments have reported events in excess of the
expected background, which would be compatible with a WIMP mass of a few
GeV. Also the CDMS-II experiment \cite{Ahmed:2009zw} has reported two
events, which could be explained by a WIMP mass of $\gsim 10$~GeV (or
background). On the other hand, exclusion limits from the Xenon10
\cite{Angle:2007uj}, Xenon100 \cite{Aprile:2010um} and CDMS-Si
\cite{Akerib:2005kh} experiments set upper bounds on the spin
independent detection cross sections for this mass range of a WIMP,
which seem incompatible with the reported hints for a signal.

In any case, it is important to know whether specific models for dark
matter with a WIMP mass of a few GeV can produce direct detection
cross sections compatible with the reported excesses, and/or whether
regions in the parameter space of such models can be tested by present
and future exclusion limits.

Supersymmetric (Susy) extensions of the Standard Model are popular,
amongst others, since they predict naturally (for unbroken R-parity, and
for a neutral Lightest Supersymmetric Particle (LSP)) a candidate for
dark matter, with a relic density compatible with WMAP constraints
\cite{Komatsu:2008hk}. Within the Minimal Supersymmetric extension of
the Standard Model (MSSM) four neutral fermions (neutralinos $\chi_i^0$,
$i=1\dots 4$) exist, which are composed of the bino (superpartner of the
$U(1)_Y$ gauge boson), the wino (superpartner of the $W_\mu^3$ gauge
boson) and two higgsinos (superpartners of neutral Higgs bosons). These
states mix, and the lightest neutralino $\chi_1^0$, which is the
lightest eigenvalue of the $4\times 4$ mass matrix, will be the LSP
(leaving aside the possibility of a sneutrino LSP).

Often the LSP is dominantly bino-like, whose mass $m_{\chi_1^0}$ is
approximately given by the soft Susy breaking gaugino mass $\sim M_1$.
Assuming unification of the three gaugino masses for the bino ($M_1$),
the winos ($M_2$) and the gluino ($M_3$) at the scale of Grand
Unification, $M_1$ is naturally the smallest among these mass terms at
the electroweak scale. However, due to the lower bound of $\sim 100$~GeV
on $M_2$ from the lower bound on chargino masses, the assumption of
unification of the three gaugino masses implies $M_1 \gsim 50$~GeV and a
similar lower bound on the mass of the LSP.

The assumption of unification of the three gaugino masses can be
dropped, however, in that case $M_1$ and hence the LSP mass
$m_{\chi_1^0}$ can be arbitrarily small. Then, on the other hand it can
become difficult to satisfy the WMAP constraint on the dark matter relic
density i.e., to reduce the dark matter relic density after the Big Bang
to an acceptable value compatible with this constraint. To this end,
dark matter annihilation processes have to be sufficiently effective.
For a LSP mass $\gsim 50$~GeV the following pair annihilation processes
can be relevant: exchange of Susy partners of fermions (sfermions, in
particular sleptons) in the t-channel, and Z-exchange or Higgs-exchange
in the s-channel (if the LSP has a sufficiently large higgsino
component). In addition, neutralinos can co-annihilate with other
sparticles if they have similar masses, but co-annihilation processes
will not be relevant for a light LSP as considered here. In the MSSM,
sufficiently effective dark matter annihilation processes impose
constraints on a light LSP:

Considering LSP annihilation via slepton exchange in the t-channel, a
lower bound $m_{\chi_1^0}\gsim 18$~GeV was derived in 
\cite{Hooper:2002nq,Belanger:2002nr} from the lower bound of $\sim
100$~GeV on the slepton masses. (Relaxing this bound to $\gsim 80$~GeV
for stau masses, one obtains $m_{\chi_1^0}\gsim 13$~GeV
\cite{Hooper:2002nq,Belanger:2002nr,Dreiner:2009ic}, unless the  LSP
mass is very small corresponding to hot dark matter
\cite{Dreiner:2009ic}.) Allowing for LSP annihilation via CP-odd Higgs
($A$) exchange in the s-channel, a lower limit $m_{\chi_1^0}\gsim 6$~GeV
was given in \cite{Bottino:2002ry,Bottino:2003cz,Belanger:2003wb,
Bottino:2008mf} from $m_A \gsim 90$~GeV for large values of $\tan\beta
\gsim 25$. However, as noted in \cite{Feldman:2010ke}, this region of
the parameter space of the MSSM is now strongly constrained by the
bounds on $B_s \to \mu^+\mu^-$. A LSP with a mass in the $5-15$~GeV
range in the MSSM has been considered in \cite{Kuflik:2010ah} without,
however, asking for a correct relic density.

In the Next-to-Minimal Supersymmetric Standard Model (NMSSM, see
\cite{Maniatis:2009re, Ellwanger:2009dp} for recent reviews), which can
solve the $\mu$-problem of the MSSM, the Higgs and neutralino sectors
are extended by gauge singlet states. As noticed in
\cite{McElrath:2005bp, Barger:2005hb, Gunion:2005rw,Cao:2009ad}, the
mass of the LSP can be considerably smaller in the NMSSM than in the
MSSM and can still be compatible with the WMAP constraint on the relic
density. This is a consequence of a light CP-odd Higgs
boson in the spectrum (on top of the CP-odd Higgs boson of the MSSM),
which can be mostly singlet-like and which is not ruled out by
LEP-constraints. Then, sufficiently large LSP annihilation cross
sections via the exchange of this additional CP-odd Higgs boson in the
s-channel may be possible even for a light LSP with mass of a few GeV.

A light LSP in the NMSSM could be a (dominantly) singlet-like state; in
this case, however, its direct detection cross sections would be tiny.
On the other hand, as in the MSSM, a light LSP in the NMSSM can
originate from a small value of $M_1$ in which case it will be
dominantly bino-like and can have larger direct detection cross
sections. These have been estimated in \cite{Gunion:2005rw}, where also
constraints on the corresponding parameter space from B-physics,
LEP and $\Upsilon$-physics were discussed. However, the
corresponding points in the parameter space given as examples in
\cite{Gunion:2005rw} suffer from a negative effective $\mu$-parameter
(which is in conflict with the measured anomalous magnetic moment of the
muon), and not all experimental constraints considered below are taken
into account.

In view of the interest in a light LSP with a mass in the $2-20$~GeV
range, we find it appropriate to study upper bounds on its direct
detection cross sections in the NMSSM. Direct detection cross sections
in the NMSSM including WMAP constraints have been studied before
in \cite{Cerdeno:2004xw,Cerdeno:2007sn, Barger:2007nv, Belanger:2008nt,
Yokozaki:2009fu, Cao:2010fi,Demidov:2010rq}, but not for the LSP mass
range considered here. Apart from WMAP constraints, we take care of a
lengthy list of experimental constraints from B-physics (important for
large $\tan\beta$ and/or relatively light charged and CP-odd Higgs
bosons as relevant here), $\Upsilon$-physics and LEP-constraints on
neutralino production. Among the latter, OPAL limits on $e^+
e^- \to \chi^0_1 \chi^0_i$ ($i > 1$) turn out to be very important.
Since these are also relevant for the MSSM, but have hardly been
discussed before (a notable exception is \cite{Dreiner:2009ic}), we
study their consequences in some detail. For the numerical analysis we
use the code NMSSMTools \cite{ Ellwanger:2004xm,Ellwanger:2005dv}
coupled to micrOMEGAs
\cite{Belanger:2005kh,Belanger:2006is,Belanger:2008sj}. As a result we
obtain upper bounds on the spin-independent and spin-dependent
LSP-nucleon cross sections of $\sigma^{SI} \lsim 10^{-42}$~cm$^{-2}$ and
$\sigma^{SD}\lsim 4\times 10^{-40}$~cm$^{-2}$, varying somewhat with the
LSP mass in the $2-20$~GeV range. The maximal value for $\sigma^{SI}$ is
indeed near the estimate given in \cite{Gunion:2005rw}.

In the next section (2) we present the relevant parameters of the NMSSM
and their impact on the LSP cross sections. In section (3), we discuss
the relevant experimental constraints. The consequences of the OPAL
constraints on $e^+ e^- \to \chi^0_1 \chi^0_i$ ($i > 1$) on the
parameter space (implying a lower bound on $\mu_{eff}$) are estimated in
an analytic approximation, which reproduces well the full numerical
results. Section (4) is devoted to our results and conclusions.

\section{The NMSSM and the impact of its parameters on the LSP cross
sections}

In the NMSSM the $\mu$ parameter of the MSSM is replaced by a Yukawa
coupling $\lambda$ to a gauge singlet (super-) field $S$. Then, the
vacuum expectation value (vev) $s$ of the real scalar component of $S$
generates an effective $\mu$-term
\beq\label{eq:1}
\mu_{eff}=\lambda s\; .
\eeq
Most of the time one studies the NMSSM with a scale invariant
superpotential $W$ which contains, apart from the Yukawa coupling of $S$
to the MSSM-like Higgs doublet fields $H_u$ and $H_d$, a trilinear term
$\sim \frac{\kappa}{3} S^3$. Hence the Higgs mass term $\mu H_u H_d$ in
$W_{MSSM}$ is replaced by
\beq\label{eq:2}
W_{NMSSM} = \lambda S H_u H_d + \frac{\kappa}{3} S^3 + \dots\; .
\eeq
(Occasionally one considers the so-called nMSSM \cite{Cao:2009ad,
Barger:2007nv, Cao:2010fi} without the trilinear coupling $\sim
\frac{\kappa}{3} S^3$, which is replaced by a tadpole-term $\sim \xi_F
S$.) Compared to the MSSM, the gauge singlet superfield $S$ adds
additional degrees of freedom to the CP-even and CP-odd Higgs sectors as
well as to the neutralino sector. Hence the spectrum contains
\begin{itemize}
\item 3 CP-even neutral Higgs bosons $H_i$, $i=1,2,3$, which mix in
general;
\item 2 CP-odd neutral Higgs bosons $A_1$ and $A_2$;
\item one charged Higgs boson $H^\pm$;
\item five neutralinos $\chi_i^0$, $i=1\dots 5$, which are mixtures of
the bino, the neutral wino, the neutral higgsinos and the singlino;
\item two charginos which are mixtures of the charged winos and the
charged higgsinos.
\end{itemize}

Apart from the Susy generalisations of the Standard-Model-like gauge and
Yukawa couplings and the superpotential in Eq.~(\ref{eq:2}), the
Lagrangian of the NMSSM contains soft Susy breaking terms in the form of
gaugino masses $M_1$, $M_2$ and $M_3$ for the bino, the winos and the
gluino, respectively, mass terms for all scalars (squarks, sleptons,
Higgs bosons including the singlet $S$) as well as trilinear scalar
self-couplings as $\lambda A_\lambda S H_u H_d +  \frac{\kappa}{3}
A_\kappa S^3$, which reflect the trilinear couplings among the
superfields in the superpotential.

Expressions for the mass matrices for all Higgs- and neutralino states
can be found in \cite{Maniatis:2009re, Ellwanger:2009dp}; below
we confine ourselves to those which are of relevance subsequently.
Dropping the Goldstone mode, the $2 \times 2$ mass matrix for the CP-odd
Higgs bosons ${\cal M}_{P}^2$ in the basis ($A_{MSSM}, S_I$) has the
elements
\bea
{\cal M}_{P,11}^2 & = & \frac{2\, \mu_\mathrm{eff}\,
(A_\lambda+\kappa s)}{\sin 2\b} \equiv M_A^2\; , \nn\\
{\cal M}_{P,22}^2 & = & \l (A_\lambda/s+4\k ){v_u v_d} -3\k
A_\k\, s\; , \nn\\
{\cal M}_{P,12}^2 & = &\l (A_\l - 2\k s)\, v
\label{eq:3}
\eea
where $v_u$, $v_d$ denote the vevs of $H_u$, $H_d$, respectively, $v =
\sqrt{v_u^2 + v_d^2} \sim 174$~{GeV} and, as usual, $\tan\beta =
v_u/v_d$. The matrix element ${\cal M}_{P,11}^2$ would be the mass
squared of the MSSM-like CP-odd scalar $A_{MSSM}$, if the singlet sector
were absent; subsequently we will denote it simply by $M_A^2$. (This
parameter can replace the parameter $A_\lambda$.) For any (possibly
large) value of $M_A^2$, ${\cal M}_{P}^2$ can have another small
eigenvalue corresponding to an additional light CP-odd Higgs boson $A_1$
which is mostly singlet-like. This state will be relevant for the LSP
annihilation cross section below.

The mass of the charged Higgs scalar is given by
\beq\label{eq:4}
{M}_{H^\pm}^2 = 
M_A^2 + v^2 (\frac{g_2^2}{2} - \l^2)\; ;
\eeq
note that it decreases with increasing $\lambda$. As is well known, too
small values of $M_{H^\pm}$ can cause disagreements between measurements
and corresponding contributions to B-physics-observables as $b \to s
\gamma$; this will be of importance below.

Notably for large $M_A$, one of the 3 CP-even Higgs bosons will have a
mass close to $M_A$. In the MSSM, the corresponding CP-even state is
denoted by $H$, and we will maintain this denomination. The
spin-independent LSP-nucleon cross section will be dominated by the
exchange of this CP-even scalar $H$, since its couplings to down-type
quarks (particularly the strange quark) are enhanced for large values of
$\tan\beta$.

Also, the mass of the charged Higgs scalar is close to $M_A$ for large
$M_A$; then the states $H$, $A_{MSSM}$ and $H^\pm$ form a nearly
degenerate SU(2) doublet. In fact this approximate degeneracy holds down
to fairly low values of $M_A \sim 300$~GeV.

In the neutralino sector, the bino $\l_1$ and the neutral wino $\l_2^3$
mix with the neutral higgsinos $\psi_d^0, \psi_u^0$ and the singlino
$\psi_S$, and generate a symmetric $5 \times 5$ mass matrix ${\cal
M}_0$. In the basis $\psi^0 = (-i\l_1 , -i\l_2^3, \psi_d^0, \psi_u^0,
\psi_S)$,
${\cal M}_0$ reads
\beq\label{eq:5}
{\cal M}_0 =
\left( \ba{ccccc}
M_1 & 0 & -\frac{g_1 v_d}{\sqrt{2}} & \frac{g_1 v_u}{\sqrt{2}} & 0 \\
& M_2 & \frac{g_2 v_d}{\sqrt{2}} & -\frac{g_2 v_u}{\sqrt{2}} & 0 \\
& & 0 & -\mu_\mathrm{eff} & -\l v_u \\
& & & 0 & -\l v_d \\
& & & & 2 \k s
\ea \right)\; .
\eeq
It can be diagonalized by an orthogonal real matrix
$N_{ij}$ such that the physical masses $m_{\chi^0_i}$ ordered in
$|m_{\chi^0_i}|$ are real (but not necessarily positive).
Denoting the 5 eigenstates by $\chi^0_i$, we have
\beq\label{eq:6}
\chi^0_i = N_{ij} \psi^0_j\; .
\eeq

Finally, the chargino masses are described by a $2 \times 2$ mass matrix
containing $M_2$ and $\mu_{eff}$ as diagonal entries. The lower bound of
$\sim 103$~GeV on the lightest chargino implies at least the constraint
\beq\label{eq:7}
Min\{M_2, |\mu_{eff}|\} \gsim 100\ \mathrm{GeV}
\eeq
(one can choose $M_2 > 0$ by convention).

Next, we discuss the dominant contribution to the spin-independent
LSP-nucleon cross section $\sigma^{SI}$. Leaving aside scenarios with
light squark masses of $\sim 100$~GeV (which are difficult to reconcile
with Tevatron constraints), $\sigma^{SI}$ is dominated by the exchange
of CP-even Higgs bosons, which couple mostly to the strange quark sea.
Among the CP-even Higgs bosons, the coupling of the state $H$ to
down-type quarks (as the strange quark) increases with $\tan\beta$.
Hence, although its mass is generally larger than the mass of the
Standard-Model-like Higgs boson $h$, $H$-exchange provides the leading
contribution to $\sigma^{SI}$ for large values of $\tan\beta$. Then, the
dominant component of $H$ is given by $H_d$.

The dominant coupling of $H \sim H_d$ to the LSP is induced by the
bino-higgsino-Higgs vertex $\sim g_1$ and hence proportional to $g_1
N_{11}N_{13}$, where $N_{11}$ denotes the bino- and $N_{13}$ the
$\psi_d^0$-higgsino-component of the LSP. All in all one finds
\beq
\label{eq:8}
\sigma^{SI} \sim N_{11}^2 N_{13}^2 \frac{\tan^2\b}{m_H^4}\; ,
\eeq
which shows that the largest values of $\sigma^{SI}$ are obtained for a
large product $N_{11}N_{13}$, large $\tan\beta$ and low values of $m_H$.

The dominant contribution to the spin-dependent LSP-nucleon cross
section $\sigma^{SD}$ originates, as in the MSSM, from $Z$-exchange. At
first sight one could imagine that, for a light CP-odd Higgs boson
$A_1$, its exchange could also give important contributions to
$\sigma^{SD}$. However, a light $A_1$ is dominantly singlet-like and,
moreover, the coupling of its doublet component to strange quarks is
always tiny compared to the $Z$-boson coupling.

The coupling of the $Z$-boson to the LSP originates from the gauge
couplings of the higgsino components $\psi_u^0$ and $\psi_d^0$. Since no
additional free parameters intervene, the spin-dependent cross section
$\sigma^{SD}$ is proportional to
\beq
\label{eq:9}
\sigma^{SD} \sim (N_{13}^2 - N_{14}^2)^2\; .
\eeq

Finally the LSP annihilation cross section $\sigma_{ann}$ is dominated,
for the LSP mass range $2-20$~GeV under consideration, by the exchange of
a light $A_1$ in the s-channel. The dominant contribution to the $A_1
\chi_1^0 \chi_1^0$ coupling is induced by the doublet component of $A_1$
and the bino-higgsino components of $\chi_1^0$ as in the case of the $H
\chi_1^0 \chi_1^0$ coupling above; the singlet components of $A_1$ and
$\chi_1^0$ play a minor r\^ole here. In any case one has
(neglecting the finite width of $A_1$ and the velocity of $\chi_1^0$
near the freeze-out temperature)
\beq
\label{eq:10}
\sigma_{ann} \sim \frac{1}{(m_{A_1}^2 - 4\, m_{\chi_1^0}^2)^2}\; ,
\eeq
and hence $\sigma_{ann}$ can be sufficiently large for suitable values
of $m_{A_1}^2$, the lightest eigenvalue of ${\cal M}_P^2$ in
Eq.~(\ref{eq:3}).

\section{Experimental constraints on the para\-meter space}

In this section we discuss various constraints on the parameters of the
NMSSM, notably (but not exclusively) from LEP and B-physics, separately
in various subsections.

\subsection{Constraints from sparticle and Higgs searches}

As we have seen in Eq.~(\ref{eq:8}), a large spin-independent detection
cross section $\sigma^{SI}$ requires bino-components $N_{11}$ and
higgsino-components $N_{13}$ of the LSP. For small $M_1$ such that
$m_{\chi_1^0}$ is in the $2-20$~GeV range, the bino component of
$\chi_1^0$ is automatically large. However, a large higgsino component
of $\chi_1^0$ require relatively small values for $\mu_{eff}$ (below
$\sim 160$~GeV) in the mass matrix ${\cal M}_0$ in Eq.~(\ref{eq:5}).
Consequently the neutralino states $\chi_2^0$ and $\chi_3^0$ (for $M_2,
2\k s > \mu_{eff}$) are higgsino-like with masses of the order of
$\mu_{eff}$. Then, the production process $e^+ e^- \to \chi_1^0
\chi_i^0$ ($i=2,3$) was kinematically possible at LEP2, and
corresponding limits from DELPHI \cite{Abdallah:2003xe} and OPAL
\cite{Abbiendi:2003sc} have to be taken into account. 

The strongest limits come from OPAL at 208~GeV, where we can assume
100\% $Z^{*}$ branching ratios for the $\chi_i^0$ decays (see Fig.~10
in \cite{Abbiendi:2003sc}). Upper bounds on the cross section are given
in 5~GeV-wide bins of $m_{\chi_i^0}$. Since we will find $m_{\chi_3^0} -
m_{\chi_2^0} \sim 40$~GeV, the bounds apply for $\chi_2^0$ and
$\chi_3^0$ separately. For $m_{\chi_1^0} < 20$~GeV, at least one of the
$\chi_2^0$ or $\chi_3^0$ production cross sections (in association with
$\chi_1^0$) is bounded from above by $0.05$~pb.

In principle, both $Z^{*}$-exchange in the s-channel and selectron
exchange in the t-channel contribute to this cross section. However, the
interference between these channels is positive, hence the most
conservative bounds on the parameters are obtained by assuming heavy
selectrons and that $e^+ e^- \to \chi_1^0 \chi_i^0$ originates from
$Z^{*}$-exchange only. The expression for $\sigma_Z(e^+ e^- \to \chi_1^0
\chi_i^0)$ is given, e.g., in \cite{Franke:1995tf} and can be written as
\bea
\sigma_Z(e^+ e^- \to \chi_1^0\chi_i^0) &=& 
\frac{(g_1^2+g_2^2)^2}{32\pi(s-M_Z^2)^2}
\left(N_{13}N_{i3}-N_{14}N_{i4}\right)^2
\left(\frac{1}{4}-\sin^2\t_w+2 \sin^4\t_w\right)\nn \\
&&\times \frac{\sqrt{\lambda(s)}}{s} \left(E_1 E_i +
\frac{\lambda(s)}{12 s} - m_{\chi_1^0} m_{\chi_i^0}\right)
\label{eq:11}
\eea
(note the different basis for the neutralinos in \cite{Franke:1995tf})
with
\beq
\label{eq:12}
\lambda(s) = s^2+m_{\chi_1^0}^4+m_{\chi_i^0}^4-2 s \left(m_{\chi_1^0}^2
+ m_{\chi_i^0}^2\right) - 2 m_{\chi_1^0}^2 m_{\chi_i^0}^2\; .
\eeq

In order to obtain an approximate expression for the resulting
constraints on the parameters, we first neglect $m_{\chi_1^0}$
everywhere in (\ref{eq:11}). Using numerical values for the gauge
couplings, (\ref{eq:11}) simplifies to
\beq
\label{eq:13}
\sigma_Z(e^+ e^- \to \chi_1^0\chi_i^0) [\mathrm{pb}] \simeq
4.9\times 10^4 \frac{(s-m_{\chi_i^0}^2)^2}{s(s-M_Z^2)^2} 
\left(1+\frac{m_{\chi_i^0}^2}{2 s}\right)
\left(N_{13}N_{i3}-N_{14}N_{i4}\right)^2
\eeq
with $s$ and the masses in GeV. Next we look for approximations for the
relevant neutralino mixing parameters $N_{ij}$. For simplification we
assume $M_2$, $2\k s \gg |\mu_{eff}|$ such that the wino- and
singlino-sectors in the mass matrix ${\cal M}_0$ in Eq.~(\ref{eq:5})
decouple. (The wino- and singlino-components of the LSP hardly
contribute to the spin-independent cross section.) Assuming, in
addition, large $\tan\beta$ such that $v_d \ll v_u$, ${\cal M}_0$ can be
diagonalised analytically with the results (we define $u=g_1
v_u/\sqrt{2} \sim 43$~GeV and write $\mu \equiv \mu_{eff}$)
\bea
&& N_{11} \sim \frac{-1}{\sqrt{1+\frac{u^2}{\mu^2}}}\; ,
N_{13} \sim \frac{-1}{\sqrt{1+\frac{\mu^2}{u^2}}}\; ,  
N_{14} \sim 0 \nn\\
&& N_{21} \sim - N_{31} \sim 
\frac{1}{\sqrt{2}\sqrt{1+\frac{\mu^2}{u^2}}}\;,\
N_{23} \sim - N_{33} \sim
\frac{-1}{\sqrt{2}\sqrt{1+\frac{u^2}{\mu^2}}}\;, \
N_{24} \sim N_{34} \sim \frac{1}{\sqrt{2}}\; .
\label{eq:14}
\eea

Replacing these expressions into (\ref{eq:13}), using the numerical
values for $s$ and $M_Z$ in the denominator and, notably, approximating
$m_{\chi_i^0} \sim \mu$, one ends up with
\beq
\label{eq:15}
\sigma_Z(e^+ e^- \to \chi_1^0\chi_i^0) [\mathrm{pb}] \simeq
8.3\times 10^{-7} \frac{(s-\mu^2)^2 \mu^2}{u^2 + \mu^2}
\left(1+\frac{\mu^2}{2s}\right)\; ,
\eeq
where $s$, $\mu$ and $u$ are in GeV. Then the upper OPAL bound on
$\sigma_Z(e^+ e^-  \to \chi_1^0\chi_i^0)$ of $0.05$~pb becomes
a lower bound on $|\mu|$ ($\equiv \mu_{eff}$),
\beq
\label{eq:16}
|\mu_{eff}| \gsim 111\ \mathrm{GeV}\; .
\eeq

A somewhat stronger version of the OPAL bound ($\sigma_Z < 0.01$~pb) is
implemented in the default version of NMSSMTools
\cite{Ellwanger:2004xm,Ellwanger:2005dv}. We replace it by the published
value of $0.05$~pb \cite{Abbiendi:2003sc} for our numerical analysis.
From this, without any approximations, we obtain $|\mu_{eff}|
\gsim 114$~{GeV} (varying somewhat with $M_2$ and $\tan\b$) for small
values of $m_{\chi_1^0}$ in good agreement with the previous estimation.
We remark that, within the approximations used in Eqs.~(\ref{eq:14}),
this implies an upper bound on $N_{13}$ of $N_{13}^2 \lsim 0.12$.

Next, we consider constraints from the upper bound on the invisible $Z$
decay width, to which the decay $Z \to \chi_1^0 \chi_1^0$ would
contribute. From \cite{:2005ema} we obtain $\Delta \Gamma_Z^{inv} \lsim
2.0$~MeV (a value slightly above the one used in \cite{Dreiner:2009ic},
but below the value used in \cite{Gunion:2005rw}). The expression for
the contribution to $\Delta \Gamma_Z^{inv}$ from $\chi_1^0$ reads
\beq
\label{eq:17}
\Delta \Gamma_Z^{inv} = \frac{M_Z^3 G_F}{12 \sqrt{2} \pi}
\left(N_{13}^2-N_{14}^2\right)^2 \left(1-\frac{4
m_{\chi_1^0}^2}{M_Z^2}\right)^{3/2} \sim 0.165\ \mathrm{GeV}
\left(N_{13}^2-N_{14}^2\right)^2\; ,
\eeq
where the last expression holds for small $m_{\chi_1^0}$. Then the upper
bound on $\Delta \Gamma_Z^{inv}$ implies
\beq
\label{eq:18}
\left|N_{13}^2 - N_{14}^2\right| < 0.11\; .
\eeq
For large $\tan\beta$, where $N_{14}^2 \ll N_{13}^2$, this bound on
$N_{13}^2$ is very similar to the bound obtained above from the OPAL
limits. According to the numerical analysis without approximations we
find that the constraints on the parameter space from $e^+ e^- \to
\chi_1^0 \chi_i^0$ are mostly somewhat stronger than those from $\Delta
\Gamma_Z^{inv}$; from (\ref{eq:8}) and (\ref{eq:9}) it should be clear,
that these constraints are relevant for upper bounds on the
spin-independent and spin-dependent LSP-nucleon cross sections.

For the chargino masses we require a lower bound of $103$~GeV
\cite{charginos}, which implies lower limits on combinations of the
parameters $M_2$ and $\mu_{eff}$. In the neutral Higgs sector we apply
the various constraints from \cite{Schael:2006cr}. Since the lightest
CP-even Higgs boson $h$ is mostly Standard-Model-like in our case, these
constraints reduce to the well-known bound $m_h > 114$~GeV. On the other
hand the constraints from B-physics, as described below, will imply
charged Higgs masses above $\sim 200$~GeV, hence additional bounds from
direct charged Higgs production are not required.

\subsection{Constraints from B-physics}

Relevant constraints from B-physics originate from bounds on $BR(b \to s
\gamma) = (3.55 \pm 0.24 \pm 0.09) \times 10^{-4}$ \cite{hfag}, $\Delta
M_s = 17.77 \pm 0.12$~ps$^{-1}$ \cite{Abulencia:2006ze} and $\Delta M_d
= 0.507 \pm 0.005$~ps$^{-1}$ \cite{hfag}, and the branching ratios
$BR(B_s \to \mu^+ \mu^-) < 5.8 \times 10^{-8}$ \cite{:2007kv} (which was
recently improved to $< 4.3 \times 10^{-8}$ at 95\% C.L.
\cite{Morello:2009wp}) and $BR(B^+ \to \tau^+ \nu_\tau) < (1.67 \pm
0.39) \times 10^{-4}$ \cite{hfag}. These constraints are implemented in
NMSSMTools as described in \cite{Domingo:2007dx}, to which we refer for
the corresponding contributions to these observables in the NMSSM. 

It should be noted that charged Higgs boson exchange contributes to
$BR(b \to s \gamma)$ and $BR(B^+ \to \tau^+ \nu_\tau)$, hence the
corresponding limits impose lower bounds on $m_{H^\pm}$. On the other
hand, Susy diagrams also contribute to these observables which depend on
parameters like $M_2$, $\mu_{eff}$, $M_{squark}$ and $A_{top}$
\cite{Buras:2002vd}. For specific choices of these parameters (notably
not too large positive values of $A_{top}$), the charged Higgs boson
contributions can be partially cancelled. This will be relevant below,
since the spin-independent LSP-nucleon cross section (\ref{eq:8}) is
maximal for small $m_H$ and, as noted above, $m_H \sim m_{H^\pm}$.

At large $\tan\b$, the observables $\Delta M_s$, $\Delta M_d$ and
$BR(B_s \to \mu^+ \mu^-)$ can receive large contributions from a light
CP-odd Higgs boson $A_1$ \cite{Domingo:2007dx} which, in turn, plays an
important r\^ole for the LSP annihilation cross section (\ref{eq:10})
for a small LSP mass $m_{\chi_1^0}$. Again, additional Susy
contributions (box diagrams) exist, leading to a complicated combination
of constraints in the parameter space. We find that, for a small LSP
mass (light $A_1$), practically all these observables impose bounds on
various corners in the parameter space.

\subsection{Additional constraints}

On the dark matter relic density we impose the $3\,\sigma$ WMAP bound
\cite{Komatsu:2008hk}
\beq
\label{eq:19}
0.091 < \Omega h^2 < 0.129\; ,
\eeq
which requires a sufficiently large LSP annihilation rate (\ref{eq:10}).

A light CP-odd Higgs boson $A_1$ with a mass below $\sim 9.3$~GeV can
appear in radiative $\Upsilon \to A_1 \gamma$ decays, on which  CLEO
\cite{:2008hs} and BaBar \cite{Aubert:2009cp,Aubert:2009cka} have
obtained upper bounds. These can be translated into the parameter space
(couplings of $A_1$) of the NMSSM
\cite{Dermisek:2006py,Domingo:2008rr,Dermisek:2010mg} and are
implemented, together with constraints from possible $A_1 - \eta_b$
mixing effects \cite{Domingo:2008rr}, in NMSSMTools. We find that these
constraints are so strong (imposing, essentially, strong upper bounds on
the $A_1 b \bar{b}$ coupling for $m_{A_1} \lsim 10$~GeV) that it becomes
very difficult to obtain a LSP annihilation rate compatible with
(\ref{eq:19}) for $m_{\chi_1^0} \lsim 2$~GeV.

Finally we require that the Susy contributions to the anomalous magnetic
moment of the muon (see \cite{Gunion:2008dg,Domingo:2008bb}
for such contributions in the NMSSM) improve the disagreement between
the result of the E821 experiment \cite{Bennett:2006fi} and the Standard
Model; as in the MSSM, this implies a positive value for $\mu_{eff}$.

\section{Results and conclusions}

Before we turn to our results, we discuss the range of parameters used
to maximise the direct detection cross sections respecting the
experimental constraints above. First, for $\lambda$ we choose a small
value $\lambda = 0.05$ such that its negative effect on $M_{H^\pm}^2$ as
in Eq.~(\ref{eq:4}) remains negligible while a non-zero doublet
component of $A_1$ induced by the off-diagonal term in Eq.~(\ref{eq:3}).
A large value for $\kappa = 0.55$ makes the singlino (and the
singlet-like CP-even Higgs state) heavy such that perturbing mixing
effects in these sectors are avoided.

For the Susy breaking squark and slepton masses we use 1~TeV such that
sleptons hardly contribute to $e^+ e^- \to \chi_1^0 \chi_i^0$. $A_{top}$
varies from 300 to 650~GeV where $H^\pm$-induced and Susy-induced
contributions to $BR(b \to s \gamma)$ tend to cancel
\cite{Domingo:2007dx}. The Susy breaking gluino and the wino masses are
chosen as $M_3 = 350$~GeV and $M_2 = 180$~GeV, respectively. (These
parameters appear in the loop-induced flavour changing $A_1$-quark
vertices \cite{Buras:2002vd}, which should be small in order to allow
for a light $A_1$ consistent with the constraints from $BR(B_s \to \mu^+
\mu^-)$.)

Although Eq.~(\ref{eq:8}) suggests that $\sigma^{SI}$ is maximised for
very large values of $\tan\beta$, the best compromise with B-physics is
obtained for reasonable values of $\tan\beta \sim 35 - 44$. Likewise,
Eq.~(\ref{eq:14}) suggests that $\mu_{eff}$ should be as small as
possible in order to maximise $N_{11} N_{13}$, but we find that the best
compromise in parameter space is obtained for $\mu_{eff} \simeq
128$~GeV. Eq.~(\ref{eq:8}) also suggests that $\sigma^{SI}$ is maximised
for $m_H$ as small as possible. However, we recall that $m_H \sim M_A
\sim m_{H^\pm}$ and that $m_{H^\pm}$ is bounded from below by several
B-physics processes. We choose $M_A$ as an input parameter of the NMSSM
(instead of $A_\lambda$) and find the largest values of $\sigma^{SI}$
for $M_A \sim 260 - 315$~GeV, implying $m_{H} \sim 205 - 260$~GeV and
$m_{H^\pm} \sim 225 - 280$~GeV where the larger values correspond to
lower masses of $m_{\chi_1^0}$ below.

\begin{figure}[ht!]
\begin{center}
\psfig{file=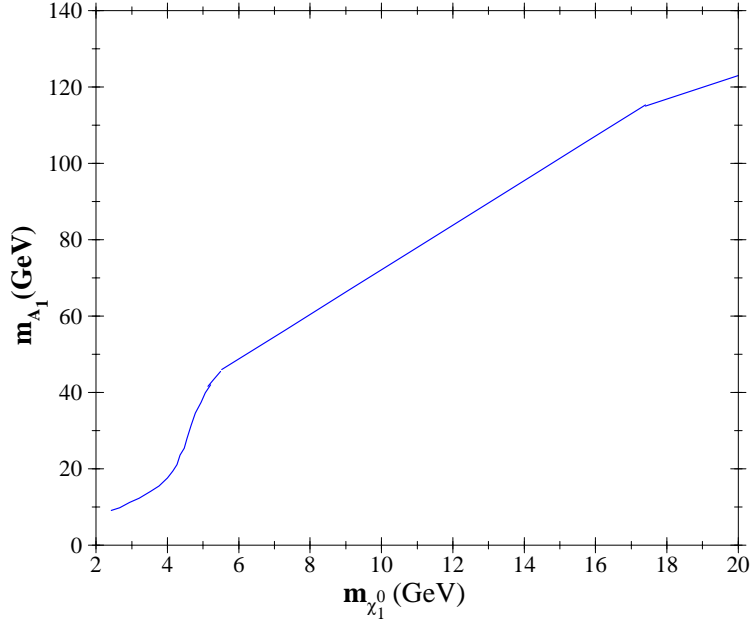, scale=0.6} 
\caption{$m_{A_1}$ as function of $m_{\chi_1^0}$ such that the relic
density of $\chi_1^0$ is in agreement with the WMAP bound
Eq.~(\ref{eq:19}).}
\label{fig:1}
\end{center}
\end{figure}

The Susy breaking bino mass term $M_1$ determines $m_{\chi_1^0}$, with
$M_1 \sim 23.5$~GeV for $m_{\chi_1^0} \sim 20$~GeV and $M_1 \sim
3.0$~GeV for $m_{\chi_1^0} \sim 2.0$~GeV. Finally $A_\kappa$ is chosen
in the range $A_\kappa \sim -14\ {\dots} -4$~GeV, which determines
$m_{A_1}$ such that $\chi_1^0$ pair annihilation via $A_1$ exchange in
the s-channel gives the correct relic density in agreement with the WMAP
bound in Eq.~(\ref{eq:19}). Due to the relatively large couplings
involved in the $\chi_1^0$ pair annihilation process, $A_1$ must
actually be off-shell and hence $m_{A_1}$ substantially larger than $2
m_{\chi_1^0}$; otherwise the relic density is too small. In
Fig.~\ref{fig:1} we show the result for $m_{A_1}$ as function of
$m_{\chi_1^0}$.

For $m_{A_1} \lsim 40$~GeV ($m_{\chi_1^0} \lsim 5$~GeV) the constraints
from $BR(B_s \to \mu^+ \mu^-)$ (where $A_1$ appears in the s-channel)
become particularly strong and require a somewhat smaller doublet
component of $A_1$. Denoting its doublet component by $\sin\t_A$, we
have $\sin\t_A \sim 0.8$ for $m_{A_1} \gsim 50$~GeV, but $\sin\t_A \sim
0.45$ for $m_{A_1} \sim 10$~GeV. We note that for $m_{A_1} \lsim 40$~GeV
the value of $A_\kappa$ has to be chosen within a precision less than
$1\%$ such that the relic density of $\chi_1^0$ is below the WMAP bound
(possibly smaller), but $m_{A_1}^2 > 0$; hence this region in the
parameter space requires considerable fine tuning. For $m_{A_1} <
10$~GeV ($m_{\chi_1^0} \lsim 2$~GeV) the constraints from CLEO and BaBar
become so strong that $\sin\t_A$ must be much smaller requiring an even
stronger fine tuning of parameters, therefore we will not consider this
range of parameters subsequently.

The components of $\chi_1^0$ (the coefficients $N_{1i}$, see
Eq.~(\ref{eq:6})) hardly change in the range $m_{\chi_1^0} = 2 - 20$~GeV
considered here, once we maximise the product $N_{11} N_{13}$ in order
to maximise $\sigma^{SI}$. We have
\bea
\label{eq:20}
&&N_{11} \sim -0.94\;,\ N_{12} \sim 0.01\dots 0.03\;,\ 
N_{13} \sim -0.32 {\dots} -0.34\;,\nn \\
&&N_{14} \sim 0.013 \dots 0.06\;,\ N_{15} \sim 0.001\; .
\eea
The masses of the mostly higgsino-like states $\chi_2^0$ and $\chi_3^0$
are $\sim 105$ and $\sim 145$~GeV, respectively, and hence as stated
before, the limits on $\sigma_Z(e^+ e^- \to \chi_1^0 \chi_i^0)$ are
relevant.

The scattering rates of $\chi_1^0$ depend somewhat on astrophysical
parameters as the escape velocity $v_{max}$ and the dark matter density
$\rho_0$ near the sun and, more importantly, on nuclear form factors
(quark matrix elements) as the pion-nucleon sigma term $\sigma_{\pi N}$
and the size of $SU(3)$ symmetry breaking parametrized by $\sigma_0$.
(The difference $\sigma_0 - \sigma_{\pi N}$ is proportional to the
strange quark matrix element.)
For the astrophysical parameters we use the default values of micrOMEGAs
$v_{max} = 544$~km/s and $\rho_0 = 0.3$~GeV/cm$^3$
\cite{Belanger:2008sj}. The default values in micrOMEGAs for
$\sigma_{\pi N}$ and $\sigma_0$ are $\sigma_{\pi N}= 55$~MeV and
$\sigma_0= 35$~MeV. 

\begin{figure}[ht!]
\begin{center}
\psfig{file=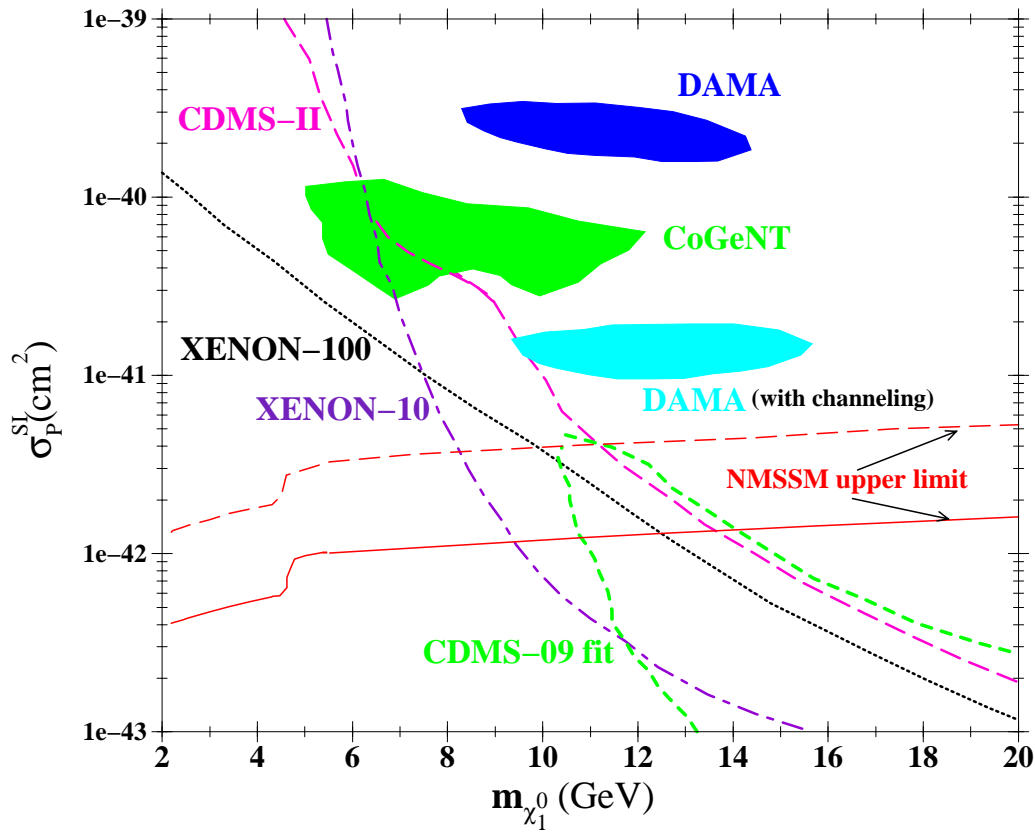, scale=0.8} 
\caption{Upper bounds on the spin-independent cross section
$\sigma^{SI}_{p}$ in the NMSSM for default values of the
strange quark content of nucleons as a full red line, and an enhanced
strange quark content of nucleons as a dashed red line. Also shown
are regions compatible with DAMA, CoGeNT and CDMS-II, and limits from
Xenon10, Xenon100 and CDMS-II as explained in the text.}
\label{fig:2}
\end{center}
\end{figure}

The corresponding results for the upper bound on the spin-independent
cross section of $\chi_1^0$ off protons $\sigma^{SI}_{p}$ in
the NMSSM are shown in Fig.~\ref{fig:2} as a function of $m_{\chi_1^0}$
as a full red line. (The spin-independent cross section off neutrons is
nearly the same.) In order to indicate the variation of this upper bound
with $\sigma_{\pi N}$ and $\sigma_0$, we show a red dashed line as the
upper bound on $\sigma^{SI}_{p}$ for $\sigma_{\pi N}= 73$~MeV and
$\sigma_0= 30$~MeV, which would correspond to a larger strange quark
matrix element and hence an increase of $\sigma^{SI}$ by a factor $\sim
3.3$.

Also shown in Fig.~\ref{fig:2} are the regions compatible with the
excesses of events reported by DAMA \cite{Bernabei:2008yi} (without
channeling (dark blue) and with channeling (light blue)), CoGeNT
\cite{Aalseth:2010vx} (light green) and a fit to the two events observed
by CDMS-II \cite{Kopp:2009qt} (denoted as CDMS-09 fit surrounded in
dashed green; these events are also compatible with background).
Exclusion limits are shown from  Xenon10 \cite{Angle:2007uj} (violet),
Xenon100 \cite{Aprile:2010um} (black) and CDMS-II \cite{Akerib:2005kh,
Ahmed:2009zw} (magenta, assuming that the two observed events originate
from background).

Fig.~\ref{fig:2} is our main result, which leads to the following
conslusions:
\begin{itemize}
\item It seems difficult to explain the excesses of events reported by
DAMA and CoGeNT within the general NMSSM (without unification
constraints on $M_1$). Hence, as stated in \cite{Gunion:2005rw},
significant modifications of parameters like a larger local dark matter
density $\rho_0$ would be required to this end. On the other hand, the
two events observed by CDMS-II (within the contour denoted as CDMS-09
fit) could be explained in the NMSSM.
\item Actual limits of Xenon10, Xenon100 and CDMS-II on spin-independent
cross sections of WIMPS in the $2 - 20$~GeV mass range test regions of
the parameter space of the NMSSM.
\end{itemize}

\begin{figure}[ht!]
\begin{center}
\psfig{file=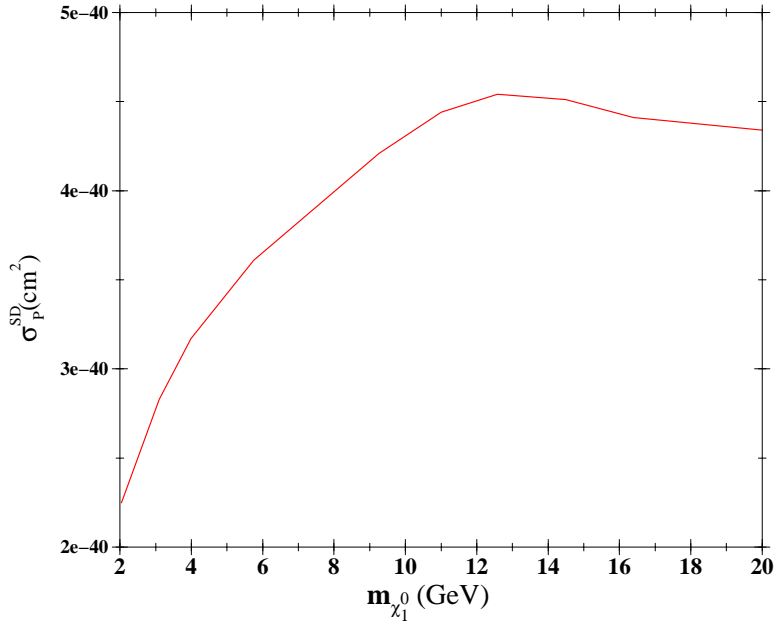, scale=0.6} 
\caption{Upper bounds on the spin-dependent cross section
$\sigma^{SD}_{p}$ in the NMSSM.}
\label{fig:3}
\end{center}
\end{figure}

For completeness we have also considered the spin-dependent cross
section $\sigma^{SD}$ in the NMSSM, which is maximal for
$\tan\beta \gsim 20$ (such that $N_{14}^2 \ll N_{13}^2$ in
Eq.~(\ref{eq:9})), large values of $M_A$ (since $m_H$ is irrelevant
here), and $\mu_{eff} \sim 121 - 129$~GeV. In Fig.~\ref{fig:3} we show
the maximum of the spin-dependent cross section off protons
$\sigma^{SD}_p$ for the same range of $m_{\chi_1^0} = 2 - 20$~GeV. Note
that $\sigma^{SD}$ originates from $Z$-exchange, hence the
spin-dependent cross section off neutrons $\sigma^{SD}_n$ is given by
$\sigma^{SD}_n \simeq 0.78 \times \sigma^{SD}_p$. The actual
experimental upper limits on $\sigma^{SD}$ are one to two orders of
magnitude larger \cite{Savage:2008er} than the upper bounds in the NMSSM
and not shown in Fig.~\ref{fig:3}.

To conclude, we have performed a detailed analysis of the parameter
space of the NMSSM for general values of $M_1$, which allows for WIMP
masses in the 2 - 20~GeV range. In contrast to the MSSM, light bino-like
WIMPs can have a relic density compatible with WMAP constraints due to a
light NMSSM-specific CP-odd Higgs state which can be exchanged in the
s-channel. Due to reported excesses of events compatible with WIMP
masses below 20~GeV, this region is of particular interest.

We have studied in detail the constraints on this region of the
parameter space of the NMSSM from LEP and B-physics, and the regions of
parameter space which give rise to maximal direct detection cross
sections while not contradicting experimental limits. The resulting
upper bounds on $\sigma^{SI} \lsim 10^{-42}$~cm$^2 = 10^{-6}$~pb make it
difficult to explain the excesses of events reported by DAMA and CoGeNT
within the NMSSM for small values of $M_1$. On the other hand, the two
events observed by CDMS-II could be explained in the NMSSM.

Notably the Xenon10 limits \cite{Angle:2007uj} on $\sigma^{SI}$ for WIMP
masses below 20~GeV start to test corresponding regions of the NMSSM
parameter space. Future results from Xenon100 could confirm the presence
of a light WIMP compatible with the NMSSM, or impose further constraints
on its parameter space.

\subsection*{Acknowledgements}

We thank A. Goudelis and Y. Mambrini for discussions. D.D. acknowledges
support from the Groupement d'Int\'er\^et Scientifique P2I.

\end{document}